\documentstyle[12pt]{article}
\begin{document}
\begin{center}
{\bf On Baxter Q-operators for Toda Chain} \\
{\bf G.P.Pronko}\\
{\it Institute for High Energy Physics , Protvino, Moscow reg.,
Russia\\
International Solvay Institute, Brussels, Belgium}
\end{center}

\vskip 1in
\begin{abstract} 
\normalsize
\par
We suggest the procedure of the construction of Baxter $Q$-operators for
Toda chain . Apart from the one-paramitric family of  $Q$-operators,
considered in our recent paper \cite {Pronko} we also give the
construction of two basic $Q$-operators and the derivation of the
functional relations for these operators. Also we have found the relation
of the basic $Q$-operators with Bloch solutions of the quantum linear
problem. 
\end{abstract}

\section {Introduction}

Long ago, in his famous papers \cite {Baxter} R.Baxter has introduced
the object, which is known now as $Q$-operator. This operator was used
initially for the solution of the eigenvalue problem of $XYZ$-spin
chain, where usual Bethe ansatz fails. Recently this operator was
intensively discussed in the series of papers \cite {BLZ} in the
connection with continuous quantum field theory. In \cite {Sklyanin1}
it was pointed out the relation of $Q$-operator with quantum
B\"aklund transformations. In \cite {Pronko} we suggested the construction
of the one-parametirc family of $Q$-operators for the most difficult
case of isotropic Heisenberg spin chain. (In spite of the obvious
simplicity of this model, the original Baxter construction fails
here.)

The existence of the one-parametric family of $Q$-operators implies the
existence of two basic solutions of Baxter equation, whose linear
combinations ( with operator coefficients ) form the one-parametric
family.

In the present paper we extend the investigation started in \cite {Pronko}
to the periodic Toda chain, the other model with rational $R$-matrix. It
turns out that apart from the construction of the one-parametric family of
$Q$-operators (section 2), in the case of Toda chain it is possible to
build also two basic $Q$-operators separately (section 3). These basic
operators satisfy to the set of the functional wronskian relations
(section 5), first established for certain field theoretical model in
\cite {BLZ}. On the one hand the wronskian relations imply the linear
independence of the basic operators, on the other hand they are the origin
for numerous fusion relations for the transfer matrix of the model. 
  
In our approach we construct the basic $Q$-operators as the trace of the
monodromy of certain $M_{n}^{(1,2)}(x)$ operators (section 3). It turns
out that these operators also permit us to construct the quantum Bloch
functions, the basis of the solutions of the quantum linear problem, which
are the eigenvectors of the monodromy matrix (section 6).

The defining relation of the $Q$-operator (Baxter equation) for the models
with rational $R$-matrix looks as follows:
\begin{equation}
t(x)Q(x)=a(x)Q(x+i)+b(x)Q(x-i),
\end{equation}
where $t(x)$ is the corresponding transfer matrix and $a(x)$ and
$b(x)$ are the c-number functions which enter into factorization of
quantum determinant of $t(x)$. In case of Toda chain the quantum
determinant is unity, therefore we can choose the normalization
$a(x)=b(x)=1$, which we shall use below.

\section {Toda Chain}

The periodic Toda Chain is the quantum system described by the Hamiltonian
\begin{equation}
H=\sum_{i=1}^{N}\left(p_i^2/2+\exp(q_{i+1}-q_{i})\right),
\end{equation}
where the canonical variables $p_{i},q_{i}$ satisfy commutation
relations
\begin{equation}
[p_{i},q_{j}]=i\delta_{ij}
\end{equation}
and periodic boundary conditions
\begin{eqnarray}
p_{i+N}&=&p_i\nonumber\\
q_{i+N}&=&q_i
\end{eqnarray}
Following Sklyanin \cite{Sklyanin2} we introduce Lax operator in
$2$-dimensional auxiliary space as follows:
\begin{eqnarray} 
L_{n}(x)=\left(\begin{array}{cc} x-p_{n}&e^{q_{n}}\\
-e^{-q_{n}}&0  \end{array}\right),
\end{eqnarray}
where x is the spectral parameter. The fundamental commutation
relations for Lax operator could be written in $R$-matrix form:
\begin{equation}
R_{12}(x-y)L_{n}^{1}(x)L_{n}^{2}(y)=L_{n}^{2}(y)L_{n}^{1}(x)R_{12}(x-
y),
\end{equation}
where indexes $1,2$ indicate different auxiliary spaces and
$R$-matrix is given by 
\begin{equation} 
R_{12}(x)=x+iP_{12},
\end{equation}
where $P$-is the operator of permutation of the auxiliary spaces.
The same intertwining relation also holds true and for the monodromy
matrix corresponding to the $L$-operator (5):
\begin{equation}
T_{ij}(x)=\left(\prod^{N}_1L_{n}(x)\right)_{ij},
\end{equation}
where the multipliers of the product is ordered from the right to the
left.

The $Q(x)$-operator we are going to construct will be given as the
trace of the monodromy $\hat Q(x)$ appropriate operators $M_{n}(x)$,
which acts in n-th quantum space and its auxiliary space, which we will
choose to be the representation space $\Gamma$ of the algebra:
\begin{equation} 
[\rho_{i},\rho^{+}_{j}]=\delta_{ij},\qquad    i,j=1,2
\end{equation}
The operator $\hat Q(x)$ will be given by the ordered product:
\begin{equation} 
\hat Q(x)=\prod_{n=1}^{N}M_{n}(x),
\end{equation}

Further we shall need to consider the product
$\left(L_{n}(x)\right)_{ij} M_{n}(x)$, which acts in the
auxiliary space $\Gamma\times C^2$ ($\Gamma$ - for $M_{n}(x)$ and $C^2$ -
is two-dimensional auxiliary space for $L_{n}(x)$). In this space it
is convenient to consider a pair of projectors $\Pi^{\pm}_{ij}$:
\begin{eqnarray} 
\Pi^{+}_{ij}&=&(\rho^{+}\rho
+1)^{-1}\rho_{i}\rho^{+}_{j}=\rho_{i}\rho^{+}_{j} (\rho^{+}\rho
+1)^{-1},\nonumber\\
\Pi^{-}_{ij}&=&(\rho^{+}\rho
+1)^{-1}\epsilon_{il}\rho^{+}_{l}\epsilon_{jm}
\rho_{m}=\epsilon_{il}\rho^{+}_{l}\epsilon_{jm}\rho_{m}(\rho^{+}\rho
+1)^{-1},
\end{eqnarray}
where
\begin{eqnarray} 
\rho^{+}\rho&=&\rho^{+}_{i}\rho_{i}\nonumber\\
\epsilon_{ij}&=&-\epsilon_{ji}, \quad \epsilon_{12}=1.
\end{eqnarray}
These projectors formally satisfy the following relations:
\begin{eqnarray} 
\Pi^{\pm}_{ik}\Pi^{\pm}_{kj}&=&\Pi^{\pm}_{ij},\nonumber\\
\Pi^{+}_{ik}\Pi^{-}_{kj}&=&0,\nonumber   \\
\Pi^{+}_{ij}+\Pi^{-}_{ij}&=&\delta_{ij}.
\end{eqnarray}

Rigorously speaking the r.h.s. of the first equation (13) in the Fock
representation has an extra term, proportional to the projector on
the
vacuum state, but, as we shall see below, this term is irrelevant in
the
present discussion.

In order to define $Q$-operator which satisfies Baxter equation  we
shall exploit Baxter's idea \cite{Baxter}, which we reformulate as
following: $M_{n}(x)$-{\it operator should satisfies the
relation}:
\begin{equation} 
\Pi^{-}_{ij}\left(L_{n}(x)\right)_{jl}M_{n}(x)\Pi^{+}_{lk}=0.
\end{equation}
If this condition is fulfilled, then
\begin{eqnarray} 
\left(L_{n}(x)\right)_{ij}
M_{n}(x)&=&\Pi^{+}_{ik}\left(M_{n}(x)\right)_{kl}
M_{n}(x)\Pi^{+}_{lj} +\nonumber\\
\Pi^{-}_{ik}\left(L(x)_{n}\right)_{kl}
M_{n}(x)\Pi^{-}_{lj}&+&\Pi^{+}_{ik}\left(L_{n}(x)\right)_{kl}
M_{n}(x)\Pi^{-}_{lj}.
\end{eqnarray}
In other words, the condition (14) guaranties that the r.h.s. of (15) in
the sense of projectors $\Pi^{\pm}$ has the triangle form and this
form will be conserved for products over $n$ due to orthogonality of
the projectors.

From (14) we obtain 
\begin{equation} 
\epsilon_{jm}\rho_{m}\left(L_{n}(x)\right)_{jk}
M_{n}(x)\rho_{k}=0.
\end{equation}
To satisfy this equation it is sufficient if
\begin{equation} 
 M_{n}(x)\rho_{k}=\left(L_{n}^{-1}(x)\right)_{kl}
\rho_{l}A_{n}(x)
\end{equation}
or
\begin{equation} 
\epsilon_{jm}\rho_{m}\left(L_{n}(x)\right)_{jk}
M_{n}(x)=B_{n}(x)\epsilon_{kl}\rho_{l},
\end{equation}
where $A_{n}(x)$ and $B_{n}(x)$ are some operators which we shall
find now. Note that the operator $L_{n}^{-1}(x)$ is given by
\begin{eqnarray} 
L_{n}^{-1}(x)=\left(\begin{array}{cc} 0&-e^{q_{n}}\\
e^{-q_{n}}&x-i-p_{n}  \end{array}\right),
\end{eqnarray}
The equation (18) could be rewritten in the following form:
\begin{equation}
\left(L_{n}^{-1}(x+i)\right)_{jk}\rho_{k}M_{n}(x)=B_{n}(x)\rho_{j}
\end{equation}
Comparing the equations (17) and (20) we conclude that they both are
satisfied provided
\begin{eqnarray}
A_{n}(x)&=&M_{n}(x-i),\nonumber\\
B_{n}(x)&=&M_{n}(x+i).
\end{eqnarray}
In such a way we obtain the following equation for the
$M(x)$-operator:
\begin{equation}
\left(L_{n}^{-1}(x+i)\right)_{jk}\rho_{k}M_{n}(x)=M_{n}(x+i)\rho_{j}.
\end{equation}
If the operator $M_{n}(x)$ satisfies this equation, the product
$L_{n}(x)M_{n}(x)$ takes the following form:
\begin{eqnarray} 
&\left(L_{n}(x)\right)_{ij}M_{n}(x)=\rho_{i}
M_{n}(x-i)\rho^{+}_{j}(\rho^{+}\rho+1)^{-1}&\nonumber\\
&+(\rho^{+}\rho+1)^{-1}\epsilon_{il}\rho^{+}_{l}
M_{n}(x+i)\epsilon_{jm}\rho_{m}+
\Pi^{+}_{ik}\left(L_{n}(x)\right)_{kl}
M_{n}(x)\Pi^{-}_{lj}&.
\end{eqnarray}
We do not detail the last term in (23) because, due to triangle
structure of  it r.h.s. this term will not enter into the trace of
$\hat Q(x)$.

Now our task is to solve the equation for $M_{n}(x)$-operator. The
detailed investigation of the equation (22) shows that the usual Fock
representation for (9) does not fit for our purpose, therefore we
shall use less restrictive holomorphic representation.

Let the operator $\rho^{+}_{i}$ be the operator of multiplication by
the
$\alpha_{i}$, while the operator $\rho_{i}$-the operator of
differentiation with respect to $\alpha_{i}$:
\begin{eqnarray} 
\rho^{+}_{i}\psi(\alpha)&=&\alpha_{i}\psi(\alpha),\nonumber\\
\rho_{i}\psi(\alpha)&=&\frac{\partial}{\partial \alpha}\psi(\alpha).
\end{eqnarray}
The operators $\rho^{+}_{i},\rho_{i}$ are canonically conjugated for
the scalar product:
\begin{equation} 
(\psi,\phi)=\int \frac{\prod_{i=1,2}d\alpha_{i}d\bar\alpha_{i}}{(2\pi
i)^{2}} e^{-\alpha\bar\alpha}\bar\psi(\alpha)\phi(\alpha)
\end{equation}
The action of an operator in holomorphic representation is defined by
its kernel:
\begin{equation} 
\left(K\psi\right)(\alpha)=\int
d^2\mu(\beta)K(\alpha,\bar\beta)\psi(\beta),
\end{equation}
where we have denoted
\begin{equation} 
d^2\mu(\beta)=\frac{\prod_{i=1,2}d\beta_{i}d\bar\beta_{i}}
{(2\pi i)^{2}}.
\end{equation} 
Now we are ready to make the following 

{\it Statement}  The kernel $M_{n}(x,\alpha,\bar\beta)$ of the
operator $M_{n}(x)$ in
holomorphic representation has the following form:
\begin{equation}
M_{n}(x,\alpha,\bar\beta)=m_{n}(x)\frac{(\alpha\bar\beta)^{2l+ix}}{
\Gamma(2l+ix+1)},
\end{equation}
where $l$ is arbitrary parameter and the operator $m_{n}(x)$ is given
by
\begin{eqnarray}
&m_{n}(x)=\exp\left[{\pi/2(\rho_{1}^+\rho_{2}e^{q_{n}}-
\rho_{2}^+\rho_{1}e^{-q_{n}}}\right]
\left(1+i\rho_{2}^+\rho_{1}e^{-q_{n}}\right)^{i(p_{n}-x)+\rho_{1}^+
\rho_{1}}\nonumber\\
&=\left(1-i\rho_{1}^+\rho_{2}e^{q_{n}}\right)^{i(p_{n}-x)+\rho_{1}^+
\rho_{1}}\exp\left[{\pi/2(\rho_{1}^+\rho_{2}e^{q_{n}}-
\rho_{2}^+\rho_{1}e^{-q_{n}}}\right]
\end{eqnarray}
In (28) the operator $m_{n}(x)$ acts on the argument $\alpha$ of the
function $(\alpha\bar\beta)^{2l+ix}$ according to (24). The proof of
the {\it Statement} is straightforward  by direct substitution of
(28) into equation (22). 
This calculation give us also the by-product -- the
meaning of the operator $m_{n}(x)$. Apparently this operator commutes
with the operator
\begin{equation}
\hat l=\frac{1}{2}(\rho_{1}^{+}\rho_{1}+\rho_{2}^{+}\rho_{2}).
\end{equation}
If we shall fix the subspace of $\Gamma$ corresponding to the
definite eigenvalue $l$  of the operator $\hat l$ then the operator
$m_{n}(x-i(l+1/2))$ becomes Lax operator of Toda chain with auxiliary
space, corresponding to the spin $l$. In particular, the operator (5)
corresponds to $l=1/2$. Generally speaking, the $m_{n}(x-i(l+1/2))$
represents Lax operator of Toda chain in the auxiliary space $\Gamma$.
This statement could be proved by intertwining of operator (5) with 
$m_{n}(x-i(l+1/2))$.

Now, taking the ordered product of the $M_n(x)$ operators we shall
obtain the operator $\hat Q(x,l)$ whose kernel is given by
\begin{eqnarray} 
\hat Q(x,l,\alpha,\bar\beta)&=&\int
\prod_{i=1}^{N-1}d^2\mu(\gamma_{i})
M_{N}(x,l,\alpha,\bar\gamma_{N-1} )
M_{N-1}(x,l,\gamma_{N-1},\bar\gamma_{N-2})\nonumber\\
\cdots&\times&M_{2}(x,l,\gamma_{2},\bar\gamma_{1})
M_{1}(x,l,\gamma_{1},\bar\beta).
\end{eqnarray}
Due to triangle (in the sense of projectors $\Pi^{\pm}$ ) structure of the
r.h.s. of (23) we obtain the following rule of multiplication of the
monodromy matrix $T(x)$ on operator $\hat Q(x)$:
\begin{eqnarray} 
&\left(T(x)\right)_{ij}\hat Q(x,l,\alpha,\bar\beta)=
(x+\frac{i}{2})^{N}\rho_{i}\hat
Q(x-i,l,\alpha,\bar\beta)\rho^{+}_{j}(\rho^{+}\rho+1)^{-1}&\nonumber\\
&(x-\frac{i}{2})^{N}(\rho^{+}\rho+1)^{-1}\epsilon_{im}\rho^{+}_{m}\hat
Q(x+i,l,\alpha,\bar\beta)\epsilon_{jk}\rho_{k}+\Pi^{+}_{im}
\bigl(\cdots\bigr)_{mk}\Pi^{-}_{kj},&
\end{eqnarray}
where we omitted the explicit expression of the last term by obvious
reason.

To proceed further we need to remind the definition of trace of an
operator in holomorphic representation. If the operator is given by its
kernel $F(\alpha,\bar \beta)$ then, (see e.g. \cite{Berezin})
\begin{equation} 
Tr F= \int d^2\mu(\alpha) F(\alpha,\bar\alpha),
\end{equation}
where the measure was defined in (27). Now we can perform the trace
operation for both sides of (32 )over the holomorphic variables and over
$i,j$ indexes, corresponding to the auxiliary $2$-dimensional space of
$T(x)$. The result is the desired Baxter equation:
\begin{equation} 
t(x)Q(x,l)=Q(x-i,l)+Q(x+i,l),
\end{equation}
where, according to (33) 
\begin{equation} 
Q(x,l)=\int d^2\mu(\alpha)\hat Q(x,l,\alpha,\bar\alpha).
\end{equation}
Note, that the trace of $\hat Q$ exists due to the exponential factor in
holomorphic measure (27) and has cyclic property, therefore $Q(x,l)$ is
invariant under cyclic permutation of the quantum variables.
Acting as above  we can also consider right multiplication
$M_{n}(x)L_{n}(x)$ to obtain
\begin{equation} 
Q(x,l)t(x)=Q(x-i,l)+Q(x+i,l).
\end{equation}
We shall not consider here the derivations of the intertwining relations
for $\hat Q(x,l)$ for different values of $x$ and $l$ and for $\hat
Q(x,l)$ and $T_{ij}(y)$. This may be done in the same way as in \cite
{Pronko} and these relations imply the following commutation relations:  
\begin{eqnarray}
\left[Q(x,l),Q(y,m)\right]&=&0\nonumber\\
\left[t(x),Q(y,l)\right]&=&0
\end{eqnarray}

In such a way we have constructed the family of solutions of the Baxter
equation which are parametrized by the parameter $l$. We can prove that
this family may be considered as a linear combinations of two basic
solutions with operator coefficients. Here arises the question - is it
possible to construct these basic operators separately. The answer is
positive and now we shall show how our procedure should be modified in
this case.

\section {Basic $Q$-operators for Toda Chain.}

As above, we shall look for the $Q$-operators in the form of the monodromy
of appropriate $M_{n}^{(i)}(x)$-operators, which we now supply with the
index $i=1,2$  and which act in $n$-th quantum space. The auxiliary space
$\Gamma$ now will be the representation space of one Heisenberg algebra,
instead of (9):
\begin{equation}
[\rho, \rho^+]=1.
\end{equation}
The product $(L_{n}(x))_{ij}M_{n}^{(i)}(x)$ is an operator in $n$-th
quantum space and in auxiliary space which is tensor product $\Gamma\times
C^2$. In this auxiliary space we shall introduce new projectors :
\begin{eqnarray}
\Pi_{ij}^{+}=\left(\begin{array}{c} 1\\ 
\rho \end{array}\right) \frac{1}{\rho^+\rho+1}(1,\rho^+),\nonumber\\
\Pi_{ij}^{-}=\left(\begin{array}{c} -\rho^+\\ 
1 \end{array}\right) \frac{1}{\rho^+\rho+2}(-\rho,1)
\end{eqnarray}
The defining equations for the operators $M_{n}^{(i)}$ ( the analogies of
eq. (14) ) are 
\begin{eqnarray}
\Pi_{ik}^-\left(L_{n}(x)\right)_{kl}M_{n}^{(1)}(x)\Pi_{lj}^+&=&0,
\nonumber\\
\Pi_{ik}^+\left(L_{n}(x)\right)_{kl}M_{n}^{(2)}(x)\Pi_{lj}^-&=&0.
\end{eqnarray}

The solutions of these equations we again will present as the kernels of
the corresponding operators in holomorphic representation of the algebra
(38):
\begin{eqnarray}
M_{n}^{(1)}(x, \alpha,\bar\beta)&=&\exp(-i\bar\beta
e^{q_{n}})\frac{e^{-\pi
x/2}}{\Gamma(-i(x-p_{n})+1)}\exp(i\alpha e^{-q_{n}}),\nonumber\\ 
M_{n}^{(2)}(x, \alpha,\bar\beta)&=&\exp(-i\alpha e^{-q_{n}})e^{-\pi
x/2}e^{(x-p_{n})}\nonumber\\
&&\qquad\times\Gamma(-i(x-p_{n})) \exp(i\bar\beta e^{q_{n}}).
\end{eqnarray}
For the right multiplication by $L_{n}(x)$ these operators automatically
satisfy the following equations:
\begin{eqnarray}
\Pi_{ik}^+M_{n}^{(1)}(x)\left(L_{n}(x)\right)_{kl}\Pi_{lj}^-&=&0,
\nonumber\\
\Pi_{ik}^-M_{n}^{(2)}(x)\left(L_{n}(x)\right)_{kl}\Pi_{lj}^+&=&0.
\end{eqnarray}
The full multiplication rules for the operators $M_{n}^{i}(x)$ and
$L_{n}(x)$ have the following form for left multiplication:
\begin{eqnarray}
&&\left(L_{n}(x)\right)_{ij}M_{n}^{(1)}(x)=\left(\begin{array}{c} 1\\ 
\rho\end{array}\right)_{i}M_{n}^{(1)}(x-i)\frac{1}{\rho^+\rho+1}(1,\rho^+)
_{j}\nonumber\\
&+&\left(\begin{array}{c} -\rho^+\\ 1 \end{array}\right)_{i}
\frac{1}{\rho^+\rho+2}M_{n}^{(1)}(x+i)(-\rho,1)_{j}+\Pi_{ik}^+
\left(L_{n}(x)\right)_{kl}M_{n}^{(1)}(x)\Pi_{lj}^-\nonumber\\
&&\left(L_{n}(x)\right)_{ij}M_{n}^{(2)}(x)=\left(\begin{array}{c} 1\\ 
\rho \end{array}\right)_{i}
\frac{1}{\rho^+\rho+1}M_{n}^{(2)}(x+i)(1,\rho^{+})_{j}\\
&+&\left(\begin{array}{c} -\rho^+\\ 
1\end{array}\right)_{i}M_{n}^{(2)}(x-i)
\frac{1}{\rho^+\rho+2}(-\rho,1)_{j}+\Pi_{ik}^-\left(L_{n}(x)\right)_{kl}M_
{n}^{(2)}(x)\Pi_{lj}^+\nonumber\\
\nonumber
\end{eqnarray}
and for right multiplication:
\begin{eqnarray} 
&&M_{n}^{(1)}(x)\left(L_{n}(x)\right)_{ij}=\left(\begin{array}{c} 1\\ 
\rho\end{array}\right)_{i}\frac{1}{\rho^+\rho+1}M_{n}^{(1)}(x-i)(1,\rho^+)
_{j}\\
&+&\left(\begin{array}{c} -\rho^+\\ 1 \end{array}\right)_{i}
M_{n}^{(1)}(x+i)\frac{1}{\rho^+\rho+2}(-\rho,1)_{j}+\Pi_{ik}^-
\left(L_{n}(x)\right)_{kl}M_{n}^{(1)}(x)\Pi_{lj}^+\nonumber\\
&&M_{n}^{(2)}(x)\left(L_{n}(x)\right)_{ij}=\left(\begin{array}{c} 1\\ 
\rho \end{array}\right)_{i}
M_{n}^{(2)}(x+i)\frac{1}{\rho^+\rho+1}(1,\rho^{+})_{j}\\
&+&\left(\begin{array}{c} -\rho^+\\ 
1\end{array}\right)_{i}\frac{1}{\rho^+\rho+2}M_{n}^{(2)}(x-i)
(-\rho,1)_{j}+\Pi_{ik}^+\left(L_{n}(x)\right)_{kl}M_
{n}^{(2)}(x)\Pi_{lj}^-\nonumber\\
\nonumber
\end{eqnarray}

These relations guaranty that the traces of the monodromies, corresponding
to both operators $M_{n}^{(i)}(x)$ satisfy Baxter equations:
\begin{eqnarray}
t(x)Q^{(i)}(x)&=&Q^{(i)}(x+i)+Q^{(i)}(x-i)\nonumber\\
Q^{(i)}(x)t(x)&=&Q^{(i)}(x+i)+Q^{(i)}(x-i)
\end{eqnarray}

We shall conclude this section with the calculation of the operators
$Q^{i}(x)$ for the simplest case of one quantum degree of freedom. In this
case from (33) we easily obtain
\begin{eqnarray}
Q^{(1)}(x)&=&\int \frac{d\alpha d \bar \alpha}{2\pi i}
e^{-\alpha \bar
\alpha} M^{1}(x,\alpha,\bar\alpha)
=\sum_{n=0}\frac{e^{-\pi x/2}}{n!}
e^{-qn}\frac{1}{\Gamma(-i(x-p)+1)}e^{qn}\nonumber\\
&=&\sum_{n=0}\frac{e^{-\pi x/2}}
{n!\Gamma(-i(x-p)+n+1)}=e^{-\pi x/2}I_{-i(x-p)}(2),
\end{eqnarray}
where $I_{\nu}(x)$ is the modified Bessel function. The analogues
calculations for the second $Q$-operator gives:
\begin{eqnarray}
Q^{(2)}(x)&=&-e^{-\pi x/2}\frac{\pi e^{\pi (x-p)}}{\sin{\pi i(x-p)}}
\sum_{n=0}\frac{1}{n!\Gamma(i(x-p)+n+1)}\nonumber\\
&&=-e^{-\pi x/2}\frac{\pi e^{\pi (x-p)}}{\sin{\pi i(x-p)}}I_{i(x-p)}(2).
\end{eqnarray}
These two expressions could be compared with the results of \cite
{Pas.God}.

\section{Intertwining Relations.}

In this section we shall consider the set of intertwining relations among
$L_{n}(x)$-operator and $M_{n}^{(i)}(x)$-operators which will imply the
mutual commutativity of transfer matrix and  $Q^{(i)}(x)$. Let us start
with the simplest relation 
\begin{equation}
R^{(i)}_{kl}(x-y)\left(L_{n}(x)\right)_{lm}M_{n}^{(i)}(y)=M_{n}^{(i)}(y)
\left(L_{n}(x)\right)_{kl}R^{(i)}_{lm}(x-y)
\end{equation}
From eq. (40) follows that for $x=y$ the $R^{(i)}$-matrixes become the
corresponding projectors - $\Pi^{-}$ for $i=1$ and $\Pi^{+}$ for $i=2$.
Making use of these properties we easily obtain:
\begin{eqnarray}
R^{(1)}_{kl}(x-y)=\left(\begin{array}{cc} x-y+i\rho^+\rho&-i\rho^+\\
-i\rho&i  \end{array}\right)\\
\nonumber\\
R^{(2)}_{kl}(x-y)=\left(\begin{array}{cc} i&i\rho^+\\
i\rho&x-y+i+i\rho^+\rho  \end{array}\right)
\end{eqnarray}

Next relation which we shall consider is 
\begin{equation}
M_{n}^{(1)}(x,\rho)M_{n}^{(2)}(y,\tau)R^{12}(x-y)=R^{12}(x-y)M_{n}^{(2)}(y
,\tau)M_{n}^{(1)}(x,\rho),
\end{equation}
where both $M$-operators act in different auxiliary spaces $\Gamma^{(i)}$
and mutual quantum space. The $R$-matrix acts in the tensor product of
auxiliary spaces $\Gamma^{(1)}\times \Gamma^{(2)}$. In (52) we have
denoted the operators which act in the auxiliary space $\Gamma^{(1)}$ as
$\rho,\rho^+$ and operators in $\Gamma^{(2)}$ as $\tau,\tau^+$. From
explicit expressions for $M$-operators (41) follows that 
\begin{eqnarray}
(\rho+\tau)M_{n}^{(1)}(x,\rho)M_{n}^{(2)}(y,\tau)&=&0,\nonumber\\
M_{n}^{(2)}(y,\tau)M_{n}^{(1)}(x,\rho)(\rho^{+}+\tau^{+})&=&0.
\end{eqnarray}
These relations mean that the products of the $M$-operators are  triangle 
operators in the  $\Gamma^{(1)}\times \Gamma^{(2)}$ and , as a result the
$R$-matrix satisfy the following equations:
\begin{eqnarray}
&(\rho+\tau)R^{12}(x)=0\nonumber\\
&R^{12}(x)(\rho^{+}+\tau^{+})=0.
\end{eqnarray}
The corollary of (54) is that the kernel of $R$-matrix in holomorphic
representation depends only on one variable:
\begin{equation}
R^{12}(x,\alpha,\bar\beta;\gamma,\bar\delta)=f(x,(\alpha-\gamma)(\bar\beta
-\bar\delta)),
\end{equation} 
where the variables $\alpha,\bar\beta$ refer to the operators
$\rho,\rho^+$ and variables $\gamma,\bar\delta$ to the operators
$\tau,\tau^+$. Taking (55) into account we can write the intertwining
relation (52) in holomorphic representation:
\begin{eqnarray}
&\int d\mu(\beta')d\mu(\delta')M_{n}^{(1)}(x,\alpha,\bar\beta')
M_{n}^{(2)}(y,\gamma,\bar\delta')f(x-y,(\beta'-\delta')(\bar\beta-
\bar\delta))=\nonumber\\
&\int d\mu(\alpha')d\mu(\gamma')f(x-y,(\alpha-\gamma)(\bar\alpha'-
\bar\gamma'))M_{n}^{(2)}(y,\gamma',\bar\delta)M_{n}^{(1)}(x,\alpha',
\bar\beta),
\end{eqnarray}
where
\begin{equation}
d\mu(\alpha)=\frac{d\alpha d\bar\alpha}{2\pi i}e^{-\alpha\bar\alpha}.
\end{equation}
To simplify this equation let us introduce the new external variables:
\begin{eqnarray}
\xi_{1}&=\frac{1}{\sqrt{2}}(\alpha+\gamma),\qquad\xi_{1}'&=\frac{1}
{\sqrt{2}}(\beta+\delta),\nonumber\\
\xi_{2}&=\frac{1}{\sqrt{2}}(\alpha-\gamma),\qquad\xi_{2}'&=\frac{1}
{\sqrt{2}}(\beta-\delta)
\end{eqnarray}
and new integration variables for l.h.s. (r.h.s.) integral:
\begin{eqnarray}
\xi''_{1}&=\frac{1}{\sqrt{2}}(\beta'+\delta')\quad&\left(\xi''_{1}=\frac{1
}{\sqrt{2}}(\alpha'+\gamma')\right)\nonumber\\
\xi''_{2}&=\frac{1}{\sqrt{2}}(\beta'-\delta')\quad&\left(\xi''_{2}=\frac{1
}{\sqrt{2}}(\alpha'-\gamma')\right).
\end{eqnarray}
Apparently, due to the structure of $M^{(i)}$-operators and $R$-matrix,
both sides of (56) depend only on the variables $\xi_{2},\bar\xi'_{2}$ and
integration over $\xi''_{1}$ becomes trivial, resulting in elimination of
these variables in the integrands. Further, representing the function
$f(x,2\xi''\bar\xi')$ as
\begin{equation}
f(x,2\xi''\bar\xi')=\sum_{n=0}C_{n}(x)\frac{(2\xi''\bar\xi')^{n}}{n!},
\end{equation}
we can perform the integration over $\xi''_{2}$ and, comparing similar
terms in both sides of (56), conclude that 
\begin{equation}
C_{n}(x)=\frac{1}{\Gamma(-ix+n+1)}.
\end{equation}
Therefore $R$-matrix in (52) has the following form in holomorphic
representation
\begin{equation}
R^{12}(x,\alpha,\bar\beta;\gamma,\bar\delta)=\sum_{n=0}\frac{\left((\alpha
-\gamma)(\bar\beta-\bar\delta)\right)^n}{n!\Gamma(-ix+n+1)}.
\end{equation}
As the operator in the space $\Gamma^{(1)}\times \Gamma^{(2)}$ the
$R$-matrix (62) is pathological because its kernel depends only on part of
holomorphic variables. In other words it contains the projector $\pi$ on
the subspace of  $\Gamma^{(1)}\times \Gamma^{(2)}$ which is formed by the
functions depending on the difference of variables. This property may be
an obstacle in the derivation  of the  commutativity of $Q$-operators from
the intertwining relation (52). The situation is saved due to the same
pathological nature of the product of $M$-operators. Indeed, let us
consider the product 
\begin{eqnarray}
&Q^{(1)}(x)Q^{(2)}(y)=Tr_{1}\prod_{k=1}^{N}M_{k}^{(1)}(x)
Tr_{2}\prod_{k=1}^{N}M_{k}^{(2)}(y)=\nonumber\\
&=Tr_{1,2}\prod_{k=1}^{N}M_{k}^{(1)}(x)M_{k}^{(2)}(y),
\end{eqnarray}
where the indexes $1,2$ mark the corresponding auxiliary space. Due to the
property (53) we can supply each term $M_{k}^{(1)}(x)M_{k}^{(2)}(y)$ in
the last product with the projector $\pi$. The same holds true also for
the product of $Q$-operators taken in the inverse order. In such a way for 
the commutativity of $Q$-operators we need to consider only the
intertwining relations of $M$-operators projected onto the space
$\pi\left(\Gamma^{(1)}\times \Gamma^{(2)}\right)$ , where our $R$-matrix
is well defined.

Next we shall consider the intertwining relation for the
$M^{(1)}$-operators with different values of spectral parameter:
\begin{equation}
R^{(11)}(x-y)M^{(1)}(x,\rho)M^{(1)}(y,\tau)=M^{(1)}(y,\tau)M^{(1)}(x,\rho)
R^{(11)}(x-y).
\end{equation}
As above, the $R$-matrix in (64) acts in the space $\Gamma^{(1)}\times
\Gamma^{(2)}$. From explicit expression for $M^{(1)}$-operator (41) we
obtain:
\begin{equation}
\rho M^{(1)}(x,\rho)=M^{(1)}(x,\rho)ie^{-q},\qquad
-ie^{q}M^{(1)}(x,\rho)=M^{(1)}(x,\rho)\rho^+
\end{equation}
These properties of $M^{(1)}$-operator imply the following conditions on
the $R$-matrix:
\begin{equation}
\tau^+ R^{(11)}(x)=R^{(11)}(x)\rho^+,\qquad \rho
R^{(11)}(x)=R^{(11)}(x)\tau,
\end{equation}
which could be satisfied if $R^{(11)}(x)$ has the following form:
\begin{equation}
R^{(11)}(x)=P_{\rho\tau}g(x,\rho^+\tau),
\end{equation}
where $P_{\rho\tau}$ denotes the operator of permutation of $\rho\tau$
variables. Substituting (67) into relation (64) we get the equation for
the  function $g$:
\begin{equation}
g(x-y,\rho^+\tau)M^{(1)}(x,\rho)M^{(1)}(y,\tau)=M^{(1)}(y,\tau)
M^{(1)}(x,\rho)g(x-y,\rho^+\tau)
\end{equation}
Making use of the explicit form of the $M^{(1)}$-operator and the formal
power series expansion for function $g$ with respect to it second argument
we can solve this equation and find the function $g$:
\begin{equation}
g(x,\rho^+\tau)=(1+\rho^+\tau)^{-ix}
\end{equation}
and therefore
\begin{equation}
R^{(11)}(x)=P_{\rho\tau}(1+\rho^+\tau)^{-ix}.
\end{equation}
As this $R$-matrix intertwines two similar objects, it should satisfies
the Yang-Baxer equation (and it really does), but we shall not investigate
further this issue. 

The last relation which we need to discuss is the intertwining of two
$M^{(2)}$-operators:
\begin{equation}
R^{(22)}(x-y)M^{(2)}(x,\rho)M^{(2)}(y,\tau)=M^{(2)}(y,\tau)M^{(2)}(x,\rho)
R^{(22)}(x-y).
\end{equation}
The $M^{(2)}$-operators also satisfy the relations analogues to (65):
\begin{equation}
\rho M^{(2)}(x,\rho)=-ie^{-q}M^{(2)}(x,\rho),\qquad
M^{(2)}(x,\rho)ie^{q}=M^{(2)}(x,\rho)\rho^+,
\end{equation}
from where we obtain the analogue of (66):
\begin{equation}
\tau^+ R^{(22)}(x)=R^{(22)}(x)\rho^,\qquad \rho^+
R^{(22)}(x)=R^{(22)}(x)\tau^+,
\end{equation}
and therefore $R^{(22)}$ has the following form:
\begin{equation}
R^{(22)}(x)=P_{\rho\tau}h(x,\tau^+\rho).
\end{equation}
Further , acting as above we find that the unknown function $h$ does
coincide with the function $g$, resulting in the following
$R^{(22)}$-matrix:
\begin{equation}
R^{(22)}(x)=P_{\rho\tau}(1+\tau^+\rho)^{-ix}.
\end{equation}

Now we have completed the derivation of all needed intertwining relation
The main corollary of these relations is the mutual commutativity of the
transfer matrix and both $Q$-operators:
\begin{equation}
[t(x),Q^{(i)}(y)]=0, \quad [Q^{(i)}(x),Q^{(j)}(y)]=0, \quad i(j)=1,2.
\end{equation}

\section {Wronskian-type Functional Relations}

It was first pointed out in \cite {BLZ} that Baxter equation (1) which
defines the $Q$ -operator could be viewed as the finite difference
analogue
of the second order differential equation which admits two independent
solution. The linear independence of the solutions could be established
through the calculation of the wronskian corresponding to the equation. In
the previous section we have constructed two solution of Baxter equation
and now our task is to prove its linear independence i.e.  to derive the
finite difference analogue of the wronskian. To solve this problem let us
consider in the details the representation of the product (63) of two
different $Q$-operators.  In the notations of the previous section
the product of two $M$-operators which enters into the r.h.s. of (63) has
the following form:
\begin{eqnarray}
&M_{k}^{(12)}(x,y,\alpha,\bar\beta,\gamma,\bar\delta)=
M_{k}^{(1)}(x,\alpha,\bar\beta)M_{k}^{(2)}(y,\gamma,\bar\delta)=\\
&e^{-\pi(x+y)/2}e^{-i\bar\beta e^{q_{k}}}
\displaystyle\frac{1}{\Gamma(-i(x-p_{k})+1)}e^{i(\alpha-\gamma)e^{-q_{k}}}
e^{\pi(y-p_{k})}\Gamma(-i(y-p_{k}))e^{i\bar\delta e^{q_{k}}}\nonumber
\end{eqnarray}
Changing the holomorphic variables according to (58) we obtain:
\begin{eqnarray}
&M_{k}^{(12)}(x,y,\xi_{1},\xi_{2},\bar\xi'_{1},\bar\xi'_{2})=
e^{-\pi(x+y)/2}e^{-i/{\sqrt{2}}(\bar\xi'_{1}+\bar\xi'_{2})
e^{q_{k}}}\nonumber\\
&\displaystyle\frac{1}{\Gamma(-i(x-p_{k})+1)}
e^{i\sqrt{2}\xi_{2}e^{-q_{k}}}
e^{\pi(y-p_{k})}\Gamma(-i(y-p_{k}))e^{i/{\sqrt{2}}
(\bar\xi'_{1}-\bar\xi'_{2})e^{q_{k}}}
\end{eqnarray}
This equation demonstrates that the kernel of $M^{(1)}(x)M^{(2)}(y)$ does
not depends on the variable $\xi_{1}$ and for calculation of the
$Q^{(1)}(x)Q^{(2)}(y)$ the dependence of (78) on the variable
$\bar\xi'_{1}$ is irrelevant because the integration over $\xi',\bar\xi'$
in (63) results in deleting $\bar\xi'_{1}$  from (78). In such a way for
the calculation of $Q^{(1)}(x)Q^{(2)}(y)$ we can use instead of
$M_{k}^{(12)}(x,y,\xi_{1},\xi_{2},\bar\xi'_{1},\bar\xi'_{2})$ the
following reduced object:
\begin{eqnarray}
&\tilde M_{k}^{(12)}(x,y,\xi,\bar\xi')=
e^{-\pi(x+y)/2}e^{-i/{\sqrt{2}}\bar\xi'e^{q_{k}}}\nonumber\\
&\displaystyle\frac{1}{\Gamma(-i(x-p_{k})+1)}
e^{i\sqrt{2}\xi e^{-q_{k}}}
e^{\pi(y-p_{k})}\Gamma(-i(y-p_{k}))e^{-i/{\sqrt{2}}
\bar\xi'e^{q_{k}}}.
\end{eqnarray}
Note that $\tilde M^{(12)}(x,y)$ is nothing else but the kernel of
$M^{(1)}(x)M^{(2)}(y)$ on the space $\pi\left(\Gamma^{(1)}\times
\Gamma^{(2)}\right)$.
Now let use expand the exponents which contain $\xi,\bar\xi$ in the r.h.s.
of (79) and move the all the factors depending on $p_{k}$ to the right:
\begin{eqnarray}
&\tilde M_{k}^{(12)}(x,y,\xi,\bar\xi')=
e^{-\pi(x+y)}\displaystyle\sum_{n,m=0}
\frac{(i\sqrt{2}\xi)^n}{n!} (-i\bar\xi'/{\sqrt{2}})^m e^{(m-n)q_{k}}
\nonumber\\
&\displaystyle\times\sum_{l=0}^{m}\frac{(-1)^{m-l}}{l!(m-l)!}
\frac{\Gamma(-i(y-p_{k})-m+l)}{\Gamma(-i(x-p_{k})+1+n-m+l)}
e^{\pi(y-p_{k})}.
\end{eqnarray}
The summation over $l$ in (80) gives:
\begin{eqnarray}
&\displaystyle\sum_{l=0}^{m}\frac{(-1)^{m-l}}{l!(m-l)!}
\frac{\Gamma(-i(y-p_{k})-m+l)}{\Gamma(-i(x-p_{k})+1+n-m+l)}=\nonumber\\
&\displaystyle\frac{(-1)^m}{m!}\frac{\Gamma(-i(y-p_{k})-m)}
{\Gamma(-i(x-p_{k})+n+1)}
\frac{\Gamma(-i(x-y)+m+n+1)}{\Gamma(-i(x-y)+n+1)}
\end{eqnarray}
and we arrive at the following expression for the $\tilde
M_{k}^{(12)}(x,y,\xi,\bar\xi')$:
\begin{eqnarray}
&\tilde M_{k}^{(12)}(x,y,\xi,\bar\xi')=
e^{-\pi(x+y)}\displaystyle\sum_{n,m=0}
\frac{(i\sqrt{2}\xi)^n}{n!}\frac{(i\bar\xi'/{\sqrt{2}})^m}{m!}
e^{(m-n)q_{k}}\nonumber\\
&\displaystyle\times\frac{\Gamma(-i(y-p_{k})-m)}
{\Gamma(-i(x-p_{k})+n+1)}
\frac{\Gamma(-i(x-y)+m+n+1)}{\Gamma(-i(x-y)+n+1)}e^{\pi(y-p_{k})}.
\end{eqnarray}
Now let $x$ and $y$ be
\begin{eqnarray}
x=z_{+}=z+i(l+1/2),\qquad
y=z_{-}=z-i(l+1/2),
\end{eqnarray}
where $l$ is an integer (half-integer).
For these values of spectral parameters (82) takes the following form:
\begin{eqnarray}
&\tilde M_{k}^{(12)}(z_{+},z_{-},\xi,\bar\xi')=
e^{-\pi z}\displaystyle\sum_{n,m=0}
\frac{(i\sqrt{2}\xi)^n}{n!}\frac{(i\bar\xi'/{\sqrt{2}})^m}{m!}
e^{(m-n)q_{k}}\nonumber\\
&\times 
\displaystyle\frac{\Gamma(-i(z_{-}-p_{k})-m)}
{\Gamma(-i(z_{+}-p_{k})+n+1)}
\frac{\Gamma(2l+m+n+2)}{\Gamma(2l+n+2)}e^{\pi(z_{-}-p_{k})}.
\end{eqnarray}
Further we need to consider (82) for the opposite shift of spectral
parameters
\begin{eqnarray}
x=z_{-}-i\epsilon,\qquad
y=z_{+}+i\epsilon.
\end{eqnarray}
We have introduced infinitesimal $\epsilon$ in (85) to remove an ambiguity
which arises in (82) for these $x$ and $y$:
\begin{eqnarray}
&\tilde M_{k}^{(12)}(z_{-},z_{+},\xi,\bar\xi')=
e^{-\pi z}\displaystyle\sum_{n,m=0}
\frac{(i\sqrt{2}\xi)^n}{n!}\frac{(i\bar\xi'/{\sqrt{2}})^m}{m!}
e^{(m-n)q_{k}}\nonumber\\
&\displaystyle\frac{\Gamma(-i(z_{+}-p_{k})-m)}
{\Gamma(-i(z_{-}-p_{k})+n+l)}
\frac{\Gamma(-2l-2\epsilon+m+n)}{\Gamma(-2l-2\epsilon+n)}
e^{\pi(z_{+}-p_{k})}.
\end{eqnarray}
For $\epsilon \to 0$ the fraction of $\Gamma$-functions in (86) takes the
following values:
\begin{eqnarray}
\lim_{\epsilon \to 0}
\displaystyle\frac{\Gamma(-2l-2\epsilon+m+n)}{\Gamma(-2l-2\epsilon+n)}=
\left\{\begin{array}{cc}
\displaystyle\frac{\Gamma(-2l+m+n)}{\Gamma(-2l+n)}, &n,m\geq 2l+1,\\
\displaystyle (-1)^m\frac{(2l-n)!}{(2l-n-m)!}, &2l\geq n+m \geq 0,\\
\displaystyle\frac{\Gamma(n+m-2l)}{\Gamma(n-2l)},&n\geq 2l\geq m\\
0,&otherwise
\end{array}\right.
\end{eqnarray}
Apparently, the vanishing of the (87) in the fourth region manifests the
triangularity of the operator the $\tilde M_{k}^{(12)}(z_{-},z_{+})$,
therefore for the calculation of the trace of the product over $k$ of
these operators we need to consider only the part of (87), corresponding
to the first two regions. Thus, the resulting expression for the twice
reduced operator has the following form:
\begin{eqnarray}
\stackrel{\approx}M_{k}^{(12)}(z_{-},z_{+},\xi,
\bar\xi')=A_{k}(z,l,\xi,\bar\xi')+B_{k}(z,l,\xi,\bar\xi'),
\end{eqnarray}
where $A$ contains the part of the r.h.s. of (86) with the summation over
$n,m$ in the region $n,m\geq 2l+1$, $B$ contains the summation over $n,m$
in the region $2l\geq n+m \geq 0$. In other words, the degrees of
$\xi,\bar\xi'$ in $A$ and $B$ have no intersection and therefore while the
calculation of the product $Q^{(1)}(z_{-})Q^{(2)}(z_{+})$ these two parts
will multiply coherently:
\begin{eqnarray}
&Q^{(1)}(z_{-})Q^{(2)}(z_{+})=\int \prod_{k=1}^{N}d\mu(\xi_{k})
\stackrel{\approx}M_{N}^{(12)}(z_{-},z_{+},\xi_{1},\bar\xi_{N})
\nonumber\\
&\times\stackrel{\approx}M_{N-1}^{(12)}(z_{-},z_{+},\xi_{N},\bar\xi_{N-1})
\cdots \stackrel{\approx}M_{1}^{(12)}(z_{-},z_{+},\xi_{2},\bar\xi'_{1})
\nonumber\\
\nonumber\\
&=\int \prod_{k=1}^{N}d\mu(\xi_{k})A_{N}(z,l,\xi_{1},\bar\xi_{N})
A_{N-1}(z,l,\xi_{N},\bar\xi_{N-1})
\cdots A_{1}(z,l,\xi_{2},\bar\xi_{1})\nonumber\\
\nonumber\\
&+\int \prod_{k=1}^{N}d\mu(\xi_{k})B_{N}(z,l,\xi_{1},\bar\xi_{N})
B_{N-1}(z,l,\xi_{N},\bar\xi_{N-1})
\cdots B_{1}(z,l,\xi_{2},\bar\xi_{1})
\end{eqnarray}
Let us consider first $A$. For the convenience we will shift the values of
$n,m$ by $2l+1$, then
\begin{eqnarray}
&A_{k}(z,l,\xi,\bar\xi)=(\xi\bar\xi')^{2l+1}e^{-\pi z}
\displaystyle\sum_{n,m=0}
\frac{(i\sqrt{2}\xi)^n}{n!}\frac{(i\bar\xi'/{\sqrt{2}})^m}{(m+2l+1)!}
e^{(m-n)q_{k}}\nonumber\\
&\times 
\displaystyle\frac{\Gamma(-i(z_{-}-p_{k})-m)}
{\Gamma(-i(z_{+}-p_{k})+n+l)}
\frac{\Gamma(2l+m+n+2)}{\Gamma(2l+n+2)}e^{\pi(z_{-}-p_{k})}.
\end{eqnarray}
Comparing (90) with (84), we see that they differ from each other by the
factor $(\xi\bar\xi')^{2l+1}$ and shift of the factorial $m!$. This
difference may be presented as appropriate transformation of $\tilde
M_{k}^{(12)}(z_{+},z_{-},\xi,\bar\xi')$ :
\begin{equation}
A_{k}(z,l,\xi,\bar\xi')=\int d\mu(\zeta)d\mu(\zeta')g_{l}(\xi,\bar\zeta)
\tilde M_{k}^{(12)}(z_{+},z_{-},\zeta,\bar\zeta')f_{l}(\zeta',\bar\xi'),
\end{equation}
where 
\begin{equation}
g_{l}(\xi,\bar\zeta)=(\xi)^{2l+1}e^{\xi\bar\zeta},\qquad
f_{l}(\zeta,\bar\xi)=
(\bar\xi)^{2l+1}\sum_{n=0}\frac{(\zeta\bar\xi)^{n}}{(n+2l+1)!}.
\end{equation}
These two functions possess the following property:
\begin{equation}
\int d\mu(\xi)f_{l}(\zeta,\bar\xi)g_{l}(\xi,\bar\zeta')=
e^{\zeta\bar\zeta'}
\end{equation}
The r.h.s. of (93) is the $\delta$-function in holomorphic representation.
But note that
\begin{equation}
\int d\mu(\xi)g_{l}(\zeta,\bar\xi)f_{l}(\xi,\bar\zeta')=
\sum_{n=2l+1}\frac{(\zeta\bar\zeta')^n}{n!}.
\end{equation}
Taking into account (93), we immediately obtain:
\begin{eqnarray}
&\int \prod_{k=1}^{N}d\mu(\xi_{k})A_{N}(z,l,\xi_{1},\bar\xi_{N})
A_{N-1}(z,l,\xi_{N},\bar\xi_{N-1})
\cdots A_{1}(z,l,\xi_{2},\bar\xi_{1})\nonumber\\
&=Q^{(1)}(z_{+})Q^{(2)}(z_{-})
\end{eqnarray}
Our next step is the consideration of $B$ part of the
$M^{(12)}(z_{-},z_{+})$. First of all we shall remove the $\sqrt{2}$ from
its holomorphic arguments, because in the integral (89) these factors will
cancelled out. Therefore we need to consider the following expression for
$B$:
\begin{eqnarray}
B_{k}(z,l,\xi,\bar\xi')=e^{-\pi
z}\displaystyle\sum_{t=0}^{2l}\sum_{m=0}^{t}
\frac{\xi^{t-m}}{(t-m)!}\frac{\bar\xi'^m}{m!}(-1)^m
i^{t+2l+1}e^{(2m-t)q_{k}}\nonumber\\
\times\frac{(2l+m-t)!}{(2l-t)!}
\displaystyle\frac{\Gamma(-i(z-p_{k})+l-m+1/2)}
{\Gamma(-i(z-p_{k})-l+t-m+l/2)}e^{\pi(z-p_{k})}
\end{eqnarray}
We intend to compare this operator with Lax operator $L^{l}_{k}(x)$ of
Toda chain with auxiliary space corresponding to the spin $l$. As it
follows from the results of the 2-nd section, $L^{l}_{k}(x)$ could be
obtain by the reduction of the operator $m_{k}(x)$ defined in (29) to the
subspace corresponding to spin $l$. In the holomorphic representation the
kernel of $L^{l}_{k}(x)$ could be easily found using the projection:
\begin{equation}
L^{l}_{k}(x,\alpha,\bar\beta)=m_{k}(x-i(l+1/2))
\frac{(\alpha\bar\beta)^{2l}}{\Gamma(2l+1)}.
\end{equation}
(Note that here we again use two-component variables
$\alpha_{i},\beta_{i}, i=1,2$). In (97) the operator $m_{k}(x)$ should be
understood as the differential operator, acting on the projection kernel
$\frac{(\alpha\bar\beta)^{2l}}{\Gamma (2l+1)}$. For the calculation of the
r.h.s. of the (97) recall that the operator exponential function in (29)
is well defined because
\begin{equation}
[i(p-x)+l_{3}, \rho_{1}^+\rho_{2}e^q]=[i(p-x)+l_{3},
\rho_{2}^+\rho_{1}e^{-q}]=0,
\end{equation}
therefore we can expand the exponential function into formal series and
find the action of each term on the projection kernel:
\begin{eqnarray}
&m_{k}(x-i(l+1/2))\displaystyle\frac{(\alpha\bar\beta)^{2l}}
{\Gamma(2l+1)}=\\
\nonumber\\
&\displaystyle\sum_{n=0}^{\infty}\frac{(-1)^n\Gamma(i(p_{k}-x)+\rho_{1}^+
\rho_{1}-l+\frac{1}{2})}{\Gamma(i(p_{k}-x)+\rho_{1}^+\rho_{1}-l-n+\frac{1}
{2})}
\frac{(i\rho_{1}^+\rho_{2}e^{q_{k}})^n}{n!}
\frac{(\alpha_{1}\bar\beta_{2}e^{q_{k}}-
\alpha_{2}\bar\beta_{1}e^{-q_{k}})^{2l}}{\Gamma(2l+1)}.\nonumber
\end{eqnarray}
Apparently, only $2l$ terms in (99) will survive because the differential
operator $(\rho_{2})^n$ acts on the polynomial. The result has the
following form:
\begin{eqnarray}
&L^{l}_{k}(x,\alpha,\bar\beta)=\displaystyle\sum_{t=0}^{2l}\sum_{m=0}^{t}
e^{(2m-t)q_{k}}\frac{\Gamma(-i(x-p_{k})-m+l+1/2)}{\Gamma(-i(x-p_{k})-m+t-l
+1/2)}\nonumber\\
&(-1)^mi^{2l+t}\displaystyle\frac{\alpha_{1}^{2l-t+m}\alpha_{2}^{t-m}
\bar\beta_{1}^{2l-m}\bar\beta_{2}^m}{(2l-t)!(t-m)!m!}
\end{eqnarray}
This $L$-operator defines the transfer matrix of Toda chain with auxiliary
space, corresponding spin $l$:
\begin{eqnarray}
t^{l}(x)=\int
\prod_{k=1}^{N}d^2\mu(\alpha_{k})L^{l}_{N}(x,\alpha_{1},\bar\alpha_{N})
L^{l}_{N-1}(x,\alpha_{N},\bar\alpha_{N-1})\cdots
L^{l}_{1}(x,\alpha_{2},\bar\alpha_{1})
\end{eqnarray}
If in this formula we will perform the integration over one pair of the
holomorphic variables, corresponding for example
$\alpha_{1},\bar\beta_{1}$ in (100), the integrand still will be presented
in the factorized form, but with new, reduced kernel of $L$-operator:
\begin{eqnarray}
&\tilde L^{l}_{k}(x,\alpha,\bar\beta)=
\displaystyle\sum_{t=0}^{2l}\sum_{m=0}^{t}
e^{(2m-t)q_{k}}\frac{\Gamma(-i(x-p_{k})-m+l+1/2)}{\Gamma(-i(x-p_{k})-m+t-l
+1/2)}\nonumber\\
&(-1)^mi^{2l+t}\displaystyle\frac{\alpha_{2}^{t-m}\bar\beta_{2}^m}{(2l-t)!
(t-m)!m!}(2l-t+m)!
\end{eqnarray}
Comparing (102) with (96) we find that
\begin{equation}
B_{k}(z,l,\xi,\bar\xi')=\tilde L^{l}_{k}(z,\xi,\bar\xi)ie^{-\pi p_{k}}.
\end{equation}
Therefore
\begin{eqnarray}
&\int \prod_{k=1}^{N}d\mu(\xi_{k})B_{N}(z,l,\xi_{1},\bar\xi_{N})
B_{N-1}(z,l,\xi_{N},\bar\xi_{N-1})
\cdots B_{1}(z,l,\xi_{2},\bar\xi_{1})\nonumber\\
&=i^{N}t^{l}(x)e^{-\pi P},
\end{eqnarray}
where 
\begin{equation}
P=\sum_{k=0}^{N}p_{k}
\end{equation}
is the integral of motion, which commutes with $t^l(x)$. In the derivation
of (104) we have moved all the factors $e^{-\pi p_{k}}$ to the right to
form $e^{-\pi P}$. Gathering together (89), (95) and (104) we obtain the
following functional relations:
\begin{eqnarray}
&Q^{(1)}(z-i(l+\frac{1}{2}))Q^{(2)}(z+i(l+\frac{1}{2}))-
Q^{(1)}(z+i(l+\frac{1}{2}))Q^{(2)}(z-i(l+\frac{1}{2}))\nonumber\\
&=i^{N}t^{l}(x)e^{-\pi P}
\end{eqnarray}
For $l=0$ the transfer matrix turns into $1$ and we have the simplest
wronskian relation:
\begin{eqnarray}
Q^{(1)}(z-i/2)Q^{(2)}(z+i/2)-
Q^{(1)}(z+i/2)Q^{(2)}(z-i/2)
=i^{N}e^{-\pi P}
\end{eqnarray}
For the illustration of this identity the reader can use the
$Q^{(i)}$-operators for one degree of freedom (47), (48). In this simplest
case (107) reduces to the well-known identity for Bessel functions:
\begin{equation}
I_{\nu}(z)I_{-\nu+1}(z)-I_{-\nu}(z)I_{\nu-1}(z)=
-\frac{2\sin(\pi\nu)}{\pi z}
\end{equation}
The general case (106) for one degree on freedom is related to Lommel
polynomials \cite {Bateman}. 

The functional relations of the type (106) was first established for
certain field theoretical model in \cite {BLZ}. In the recent paper of the
author with Yu.Stroganov \cite {PS} we have discussed the analogues
relation for the eigenvalues of $Q$-operators in the case of isotropic
Heisenberg spin chain. Originally, since the Baxter paper\cite {Baxter}
the existence of one $Q$-operator was considered as important alternative
for Bethe ansatz. The relations (106) show the importance of the second
$Q$-operator which together with the first one give rise to the numerous
fusion relations (see e.g.\cite {BLZ},\cite {PS}).

\section {Discussion}

The approach we have considered in the present paper could be applied also
to the other with rational $R$-matrix -- the discrete self-trapping (DST)
model, considered in \cite {Sklyanin1}. The quantum determinant  of Lax
operator for this model is not unity and Baxter equation has the following
form:
\begin{equation}
t(x)Q(x)=(x-i/2)^N Q(x-i)+Q(x+i).
\end{equation}
The general properties of the $Q$-operators for DST-model are similar to
that of Toda system. The eigenvalues of one $Q$-operator are polynomial in
spectrum parameter , while the eigenvalues of the second are meromorphic
functions. In the case of Toda system the eigenvalues of $Q^{(1)}(x)$ are
entire functions, the eigenvalues of $Q^{(2)}(x)$ are meromorphic. For the
DST -model there also exist the functional relations similar to (106).

The most interesting would be the application of the formalism to the case
of XXX-spin chain. The situation here is the following. In \cite {Pronko}
we have constructed the family of $Q(x,l)$-operators similar to (31).
Moreover, from the results of \cite {PS} follows that for XXX-spin chain
there exist the basic $Q$-operators. Making use of the formalism of the
section 3, it is possible to the find the $M^{(i)}_{k}(x)$-operators for
this case, but the trace of monodromies corresponding to $M^{(i)}_{k}(x)$
diverges.  This puzzle deserves further investigation.

Another interesting point we want to discuss is the relation of our
$M^{(i)}_{k}(x)$-operators with quantum linear problem for Lax operator
(5). In classical case the linear problem is the main ingredient of
inverse scattering method, in the same time for the quantum theory it
seems to be unnecessary (see for example excellent review on the subject
\cite {Faddeev}). However, let us consider the following problem:
\begin{equation}
\psi_{n+1}(x)=L_{n}(x)\psi_{n}(x),
\end{equation}
where $L_{n}(x)$ is given in (5) and $\psi_{n}$ is two component quantum
operator. From the multiplication rules (43) we obtain:
\begin{eqnarray}
\left(L_{n}(x)\right)_{ij}M_{n}^{(1)}(x)\left(\begin{array}{c} 1\\ 
\rho\end{array}\right)_{j}=\left(\begin{array}{c} 1\\ 
\rho\end{array}\right)_{i}M_{n}^{(1)}(x.i.).
\end{eqnarray}
Now let us define the operator
\begin{eqnarray}
\left(\psi_{n}^{(1)}\right)_{i}(x)=Tr\left(\prod_{k=n}^{N}M_{k}^{(1)}(x)
\left(\begin{array}{c} 1\\ 
\rho\end{array}\right)_{i}\prod_{k=1}^{n-1}M_{k}^{(1)}(x-i)\right),
\end{eqnarray}
where the trace is taken over auxiliary space. Apparently, due to (111)
the operator (112) does satisfy the equation (110). For $n=1$, the
solution has the following form:
\begin{eqnarray}
\left(\psi_{1}^{(1)}\right)_{i}(x)=Tr\left(\prod_{k=1}^{N}M_{k}^{(1)}(x)
\left(\begin{array}{c} 1\\ 
\rho\end{array}\right)_{i}\right)=Q^{(1)}(x)\left(\begin{array}{c} 1\\ 
ie^{q_{N}}\end{array}\right)_{i}, 
\end{eqnarray}
where on the last step we have used the explicit form of
$Q^{(1)}(x)$-operator for the calculation of the trace. On the other hand
the solution (113) translated to the period $N$ by the monodromy (8), due
to (111) is given by
\begin{eqnarray}
\left(\psi_{N+1}^{(1)}\right)_{i}(x)=Tr\left(
\left(\begin{array}{c} 1\\ 
\rho\end{array}\right)_{i}\prod_{k=1}^{N}M_{k}^{(1)}(x-i)\right)=Q^{(1)}
(x-i)\left(\begin{array}{c} 1\\ 
ie^{q_{N}}\end{array}\right)_{i}.
\end{eqnarray}
In other words we obtain
\begin{equation}
T_{ij}(x)\left(
\psi_{1}^{(1)}\right)_{j}(x)=\frac{Q^{(1)}(x-i)}{Q^{(1)}(x)}\left(
\psi_{1}^{(1)}\right)_{i}(x).
\end{equation}
This equation may be understood as quantum analogue of the property of
Bloch solutions, which are the eigenvectors of the translation to the
period.

Similarly we can consider the second solution. Indeed, from the
multiplications rules (43) for the $M_{n}^{(2)}(x)$ we obtain:
\begin{eqnarray}
\left(L_{n}(x)\right)_{ij}M_{n}^{(2)}(x)\left(\begin{array}{c} -\rho^+\\ 
1 \end{array}\right)_{j}=\left(\begin{array}{c} -\rho^+\\ 
1\end{array}\right)_{i}M_{n}^{(2)}(x-i).
\end{eqnarray}
Therefore the operator
\begin{eqnarray}
\left(\psi_{n}^{(2)}\right)_{i}(x)=Tr\left(\prod_{k=n}^{N}M_{k}^{(2)}(x)
\left(\begin{array}{c} -\rho^+\\ 
1\end{array}\right)_{i}\prod_{k=1}^{n-1}M_{k}^{(2)}(x-i)\right),
\end{eqnarray}
possesses the same properties as (112). The initial value of (117) is
given by
\begin{eqnarray}
\left(\psi_{1}^{(2)}\right)_{i}(x)=Tr\left(\prod_{k=1}^{N}M_{k}^{(2)}(x)
\left(\begin{array}{c} -\rho^+\\ 
1\end{array}\right)_{i}\right)=Q^{(2)}(x)\left(\begin{array}{c}-ie^{q_{1}}
\\ 1\end{array}\right)_{i}, 
\end{eqnarray}
where we again on the last step have used the explicit form of
$M^{(2)}(x)$. As above we obtain 
\begin{equation}
T_{ij}(x)\left(
\psi_{1}^{(2)}\right)_{j}(x)=\frac{Q^{(2)}(x-i)}{Q^{(2)}(x)}\left(
\psi_{1}^{(2)}\right)_{i}(x).
\end{equation}
In such a way using $M^{(i)}_{n}(x)$-operators we succeeded in the
construction of the operators which may be interpreted as the quantum
analogues of the Bloch functions of the corresponding linear problem. 
In the classical theory of finite-zone "potentials", two Bloch solutions
of the linear problem, as the functions of spectral parameter are actually
the projections of the Backer-Akhiezer function, which is the
single-valued meromorphic functions on an hyper-elliptic surface. In
quantum case the Bloch functions (112) and (117) do not possess the
branching points (in weak sense) which is the trace of the projection in
the classical case, therefore their intimate relation is somehow hidden
and it will be very interesting to uncover this relationship.

\section {Acknowledgments}

The author is grateful to  E.Skyanin, S.Sergeev, for their interest and
discussions, This work was supported in part by ESPIRIT project NTCONGS 
and RFFI Grant 98-01-0070.

\end{document}